\documentclass[aps,prl,twocolumn,showpacs,superscriptaddress,preprintnumbers,amsmath,amssymb]{revtex4}
\usepackage{amsmath}
\usepackage{graphicx}
\usepackage{dcolumn}
\usepackage{bm}
\usepackage{float}
\usepackage{xcolor}
\begin{document}
\preprint{\textit{Preprint}}
\title{Coherent diffraction radiation of relativistic terahertz pulses from a laser-driven micro-plasma-waveguide}
\author{Longqing Yi}
\thanks{longqing@chalmers.se}
\affiliation{Department of Physics, Chalmers University of Technology, 41296 Gothenburg, Sweden}
\author{T\"unde F\"ul\"op}
\affiliation{Department of Physics, Chalmers University of Technology, 41296 Gothenburg, Sweden}

\date{\today}

\begin{abstract}
We propose a method to generate isolated relativistic terahertz (THz) pulses using a high-power laser irradiating a mirco-plasma-waveguide (MPW). When the laser pulse enters the MPW, high-charge electron bunches are produced and accelerated to $\sim$ 100 MeV by the transverse magnetic modes. A substantial part of the electron energy is transferred to THz emission through coherent diffraction radiation as the electron bunches exit the MPW. We demonstrate this process with three-dimensional particle-in-cell simulations. The frequency of the radiation is determined by the incident laser duration, and the radiated energy is found to be strongly correlated to the charge of the electron bunches, which can be controlled by the laser intensity and micro-engineering of the MPW target. Our simulations indicate that 100-mJ level relativistic-intense THz pulses with tunable frequency can be generated at existing laser facilities, and the overall efficiency reaches 1$\%$.

\end{abstract}
\maketitle

High power terahertz (THz) pulses have attracted significant attention since they can serve as a unique and versatile tool in fields ranging from biological imaging to material science \cite{Tonouchi2007, Siegel2004, Cole2001, Hoffmann2011}. In particular, at high intensities, such pulses allow manipulation of the transient states of matter, for example giving control over the electronic, spin and ionic degrees of freedom of molecules and solids \cite{Kampfrath2013}. Several methods such as two-color laser filamentation \cite{Oh2014}, optical reflection in lithium-niobate \cite{Hirori2011,Fulop2014} or organic crystals \cite{Vicario2014}, and relativistic laser irradiated plasmas \cite{Sheng2005, Gopal2013, Li2012, Liao2016, Tian2017, Chen2015, Herzer2018, Liao2019, Thiele2019}, have been developed for generation of THz pulses with electric fields above 1~MV/cm. However, scaling up such methods towards higher intensities remains challenging, thus representing an active research field.

Relativistic electron beams have also been used to produce THz radiation through a variety of mechanisms that include synchrotron radiation \cite{Carr2002}, transition radiation \cite{Ginzburg1982, Happek1991}, and diffraction radiation \cite{Dnestrovskii1959, Shibata1995}. Radiation emitted by these mechanisms is coherent if the bunch length is shorter than the radiated wavelength of interest. The radiated energy  then scales as the square of the beam charge. Previous studies have also shown that the radiation power decreases significantly with the beam divergence, and the energy radiated in a small cone near-axis would strongly benefit from a high beam energy \cite{Schroeder2004}. Therefore, choosing an electron source with desired qualities (high charge, high energy, and well-collimated) can be crucial for producing intense THz emission that is attractive to a range of applications \cite{Kampfrath2013}.

Currently available sources of relativistic electron beams are either linear accelerators or compact sources based on laser-plasma acceleration. The THz radiation energy from  linear accelerators has reached $\sim$ 600 $\mu$J/pulse \cite{Wu2013}, but such sources are expensive and large and thus can only offer limited accessibility. Laser wakefield acceleration in the nonlinear ``bubble'' regime can produce multi-GeV electron beams with small divergence ($\sim$0.1~mrad), but only small charge (1-100 pC) \cite{Leemans2014}. Self-modulated laser-wakefield acceleration can produce nano-Coulomb (nC) electron bunches \cite{Leemans2002} but typically have a temperature of a few MeV, and the beam divergence is large due to direct laser acceleration \cite{Gahn1999}. Last but not least, hot electrons that arise from laser-solid interaction can reach up to nC-$\mu$C charge, but the electron temperature is typically only a few hundreds of keVs to a few MeVs, and the divergence is usually large ($\sim40^{\circ}$) \cite{Liao2016}. Recently, THz radiation energy above millijoule (mJ) level has been reported in laser-solid interaction \cite{Liao2019}, but since a picosecond laser pulse is used, the coherent frequency range is below 1 THz, and the efficiency is  $\sim0.1\%$.


\begin{figure}[!t]
\centering
\includegraphics[width=8.5cm]{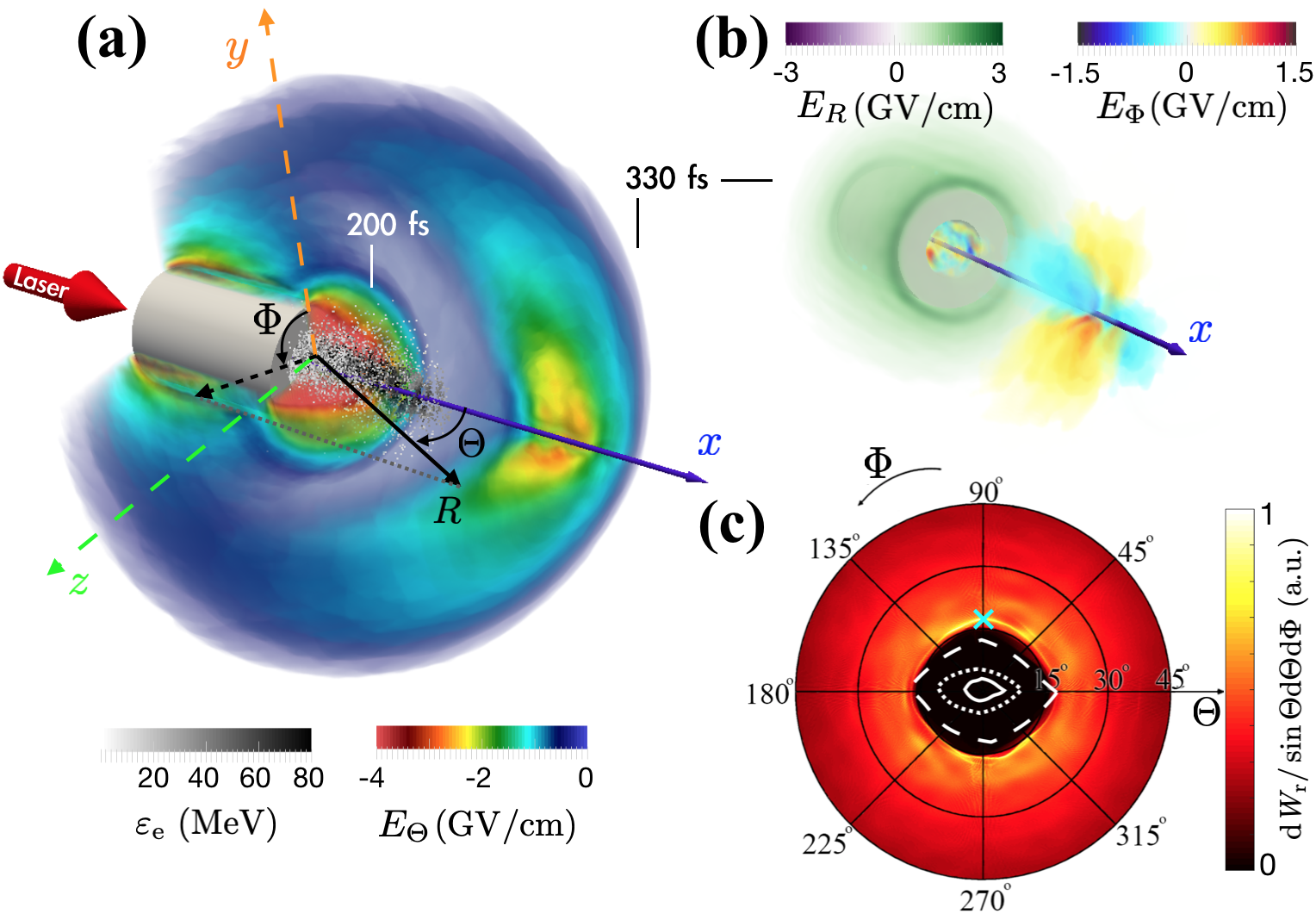}\caption{\label{f1} (Color online)  (a) Schematics of the proposed setup. A laser pulse is focused on the entrance of a MPW (white cylinder), which produces electrons and accelerates them up to $\sim$ 100 MeV. As the electron beam (black-white dots, the colour represents their energy) exits the MPW (200 fs), strong THz emission is generated by CDR. $E_\Theta$ field at 200 fs and 330 fs is shown in (a), where a quarter is removed to show the colour scale inside the radiation shell. (b)  $E_R$ and $E_\Phi$ components at 330 fs, where $R$, $\Theta$, $\Phi$ are defined in (a). The angular distribution of the radiated THz energy at 330 fs within $\Theta<45^{\circ}$ is presented in (c). The white dashed, dotted, and solid lines in (c) are contours of electron beam density at 0.1, 0.4 and 0.7 of the maximum density, respectively. The cyan cross marks the observation angle in Fig.~2.}
\vspace{-1pt}
\end{figure}

In this letter, we propose a scheme to  generate isolated THz pulses with electric fields beyond 1~GV/cm with high efficiency ($\sim1\%$). As illustrated in Fig.~1(a), an intense laser pulse is focused into a micro-plasma-waveguide (MPW), leading to electrons being extracted from the wall and accelerated by longitudinal electric fields of the transverse magnetic modes up to a few hundreds of MeVs  \cite{Bulanov1994,Yi2016_1,Yi2016_2,Yi2016_3,Gong2019}. The divergence is usually a few degrees and the duration is the same as the laser pulse. Typically the electron beam inherits the density of the plasma skin layer ($n_{\rm{beam}}\sim n_{\rm{c}}$, where $n_{\rm{c}} = m_{\rm{e}}\omega_0^2/4\pi e^2$ is the critical density, $e$ and $m_{\rm{e}}$ are the elementary charge and the electron mass, $\omega_0$ is the laser angular frequency) from which it is generated \cite{Naumova2004}. Thus, a total charge of a few tens of nC can easily be obtained with contemporary $100$-fs high-power laser systems.
Such electron source is suitable for THz generation based on coherent transition radiation and/or coherent diffraction radiation (CDR) as noted above. The simulations demonstrate that when the electron beam exits the MPW, a substantial part of the electron energy ($\sim 20\%$) is transferred to electromagnetic energy through CDR, leading to relativistically strong THz pulses up to 10-100 mJ energy.

We demonstrate our scheme using 3-dimensional (3D) particle-in-cell (PIC) simulations with the EPOCH code \cite{Arber2015}. A linearly polarised (in $y$-direction) laser with intensity 1.4$\times$ 10$^{20}$ W/cm$^{2}$ (normalised intensity $a_0 = eE_0/m_{\rm{e}}c\omega_0 = 10$, where $E_0$ is the amplitude of laser electric field and $c$ is the speed of light) is focused on the entrance of a MPW, propagating along the $x$-axis. The laser beam has a temporal Gaussian profile with FWHM duration of $\tau_0 = 54$ fs and a focal spot $w_0 = 4\lambda_0$, where $\lambda_0 = 1~\mu$m is the laser wavelength. 
The MPW has a density of $n_0 =15 n_{\rm{c}}$, the radius and length are $r_0 = 5~\mu$m and $L = 30~\mu$m, respectively. The inner surface of the MPW ($r<r_0$, where $r = \sqrt{y^2+z^2}$) has a density gradient $n(r) = n_0\exp{[-(r-r_0)^2/\sigma_0^2]}$, and the scale length is $\sigma_0 = 1~\mu$m. This leads to an effective MPW radius to be $r_{\rm{c}} = 3.35\mu$m, where $n(r_{\rm{c}}) = 1~n_{\rm{c}}$. The dimensions of the simulation box are $x\times y\times z = 100\rm{\mu m} \times 80\rm{\mu m} \times 80\rm{\mu m}$ and are sampled by $2500\times 800\times 800$ cells with 8 macro particles for electrons and 2 for C$^6+$ ions. The algorithm proposed by Cowan et al.~\cite{Cowan2013} is used to minimise the numerical dispersion.

The electric fields in the simulation with frequency below 60 THz are presented in Fig.~1(a-b), where a 35-mJ THz pulse is obtained, and the radiated power reaches 0.7 TW. To show the polarisation of the CDR, we apply spherical coordinates with the origin at the exit of MPW on the laser propagation axis $x_{\rm{c}} = 31 \mu$m, and convert the coordinates according to $R = \sqrt{(x-x_{\rm{c}})^2+y^2+z^2}$, $\Theta = \arcsin(r/R)$, and $\Phi = \arctan(z/y)$ as illustrated in Fig.~1(a).

The radiation fields are emitted simultaneously with electron propagation through the aperture at $x=x_{\rm{c}}$, mostly confined in a spherical shell. The THz emission is predominantly radially polarised in the plane determined by the observation line-of-sight and the laser propagation axis. The  polar component, $E_\Theta$, contains 99$\%$ of the radiation energy. The preference of electron distribution in the laser polarisation direction results in a small quadrupolar azimuthal electric field $E_\Phi$. The radial component $E_R$ is negligible in the radiating shell.




 


The angular distribution of THz energy in the forward direction is shown in Fig.~1(c), with white lines in the centre representing the electron beam density.
The electrons reach an cut-off energy of 100 MeV when they exit the MPW, and their total charge is 7.4 nC. The divergence of the electron beam is about $10^{\circ}$. It is slightly elongated along the laser polarisation direction.
A depleted area is observed within the electron beam, because the radiation fields add coherently and tend to cancel each other in this region. Since the electron energy is high, the CDR power is strongly peaked on the edge of the beam \cite{Carron2000}. The intensity rises sharply forming a very thin layer ($\Delta\Theta\sim1^{\circ}$) around $\Theta\sim17^{\circ}$. The radiation power in the direction perpendicular to the laser polarisation direction is higher due to the coherent sum of electric fields radiated by the elongated electron beam. 

The radiation field seen at $\Theta = 17^{\circ}$, $\Phi = 90^{\circ}$ 
is shown by the black line in Fig.~2(a). The red line shows the low-frequency component below 60 THz. The amplitude of the half-cycle THz pulse is 3 GV/cm, corresponding to a normalised amplitude of $a_{\rm{THz}} = 1.6$, reaching the relativistic intensity. Figure~2(b) shows the spectra of the radiation fields: most of the pulse energy concentrates in the desired THz frequency range of 1-10 THz. The inset in Fig.~2(b) shows the spectrum from 0 to 1000 THz for $\tau_0 = 54$ fs, with a small bump around the laser frequency at 300 THz and a peak at the double frequency 600 THz. The latter is the result of the modulation of  the electron beam  at $2\omega_0$, and can serve as an experimental signature of the CDR \cite{Schroeder2004}. 

\begin{figure}[!t]
\centering
\includegraphics[width=8.5cm]{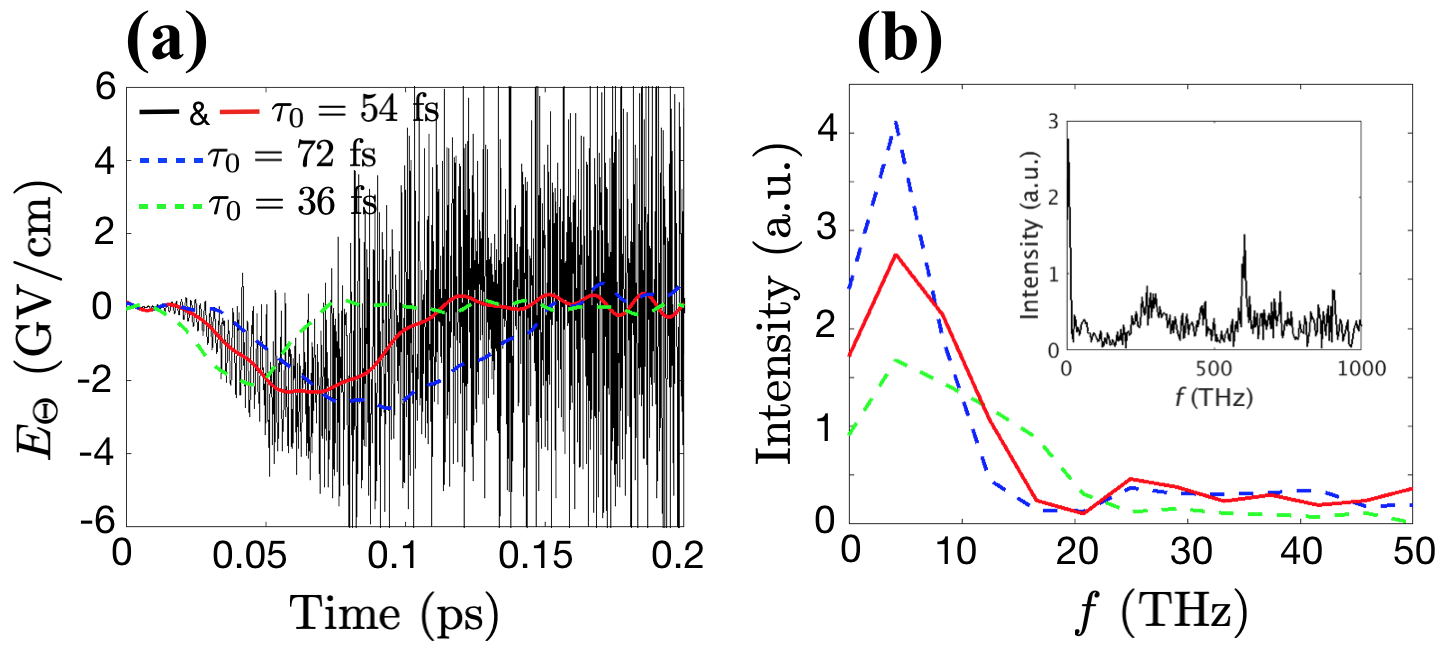}\caption{\label{f2} (Color online) The radiation fields $E_\Theta$ (a) and their spectra (b) observed at $\Theta = 17^{\circ}$ and $\Phi = 90^{\circ}$, the black and red lines represent the CDR with and without a low-pass filter (frequency below 60 THz) for laser FWHM duration $\tau_0 = 54$ fs. The green and blue dashed lines show the cases driven by $\tau_0$ equals to 36 fs and 72 fs, respectively. The inset in (b) is the full-spectrum (0-1000 THz) for $\tau_0 = 54$ fs.}
\vspace{-1pt}
\end{figure}

The duration of the THz field coincides with the laser pulse, and the cut-off frequency is determined accordingly, as shown in Fig.~2, where the  green and blue dashed lines represent the radiation field (after frequency-filtering) and spectra produced by laser pulses with $\tau_0$ of 36 fs and 72 fs, respectively. With currently available laser systems, this scheme is capable of generating relativistic pulses with frequencies ranging from near infrared to sub-THz. 

The mechanism of the electron beam generation, i.e. the electron injected into the channel (vacuum core of the MPW), is crucial for understanding the THz radiation power. The production of electron bunches at a plasma-vacuum interface when irradiated by an intense laser pulse can be attributed to the counterstreaming electrons percolating through the laser nodes, as the laser pushes the surface electrons inwards~\cite{Naumova2004}. In the MPW, the underlying physics is similar, but the mechanism that pushes the surface electrons (towards the plasma cladding), and the associated radial counterstreaming, depends on the ratio of the laser focal spot size ($w_0$) and the effective MPW radius ($r_{\rm{c}}$), which results in different injection behaviour.

To show this, we perform 2D PIC simulations of lasers having different focal spot sizes propagating in a long waveguide (240 $\mu$m). The laser and plasma parameters are the same as in the 3D simulation unless otherwise described. The resolution is 50 and 20 cells per laser wavelength in longitudinal and transverse directions, respectively. The third dimension is assumed to be 4 $\mu$m when estimating the electron number. In Fig.~3, we plot the total electron number (above 10 MeV) produced in the laser-MPW interaction against the propagation distance for different $w_0/r_{\rm{c}}$ ratios (the laser energy is fixed). 



Figure 3 shows that when $w_0/r_{\rm{c}}\geq 1$, the injection happens very fast, mainly at the entrance of the MPW. This is because the initial impact of the laser and MPW front surface is violent, which leads to strong diffracted light that pushes surface electrons into the plasma, thus results in significant counterstreaming. The electrons are more likely to be injected at this stage. In the cases where $w_0/r_{\rm{c}}<1$, the injection at the entrance is significantly reduced and the injection inside the MPW  becomes important, which is due to the interaction between waveguide modes and the MPW wall. The photon momentum $\hbar k_{\rm{T}}$, associated with the transverse wave number $k_{\rm{T}}$, pushes the surface plasma radially as the light is bouncing between the walls.

\begin{figure}[!b]
\centering
\includegraphics[width=6.0cm]{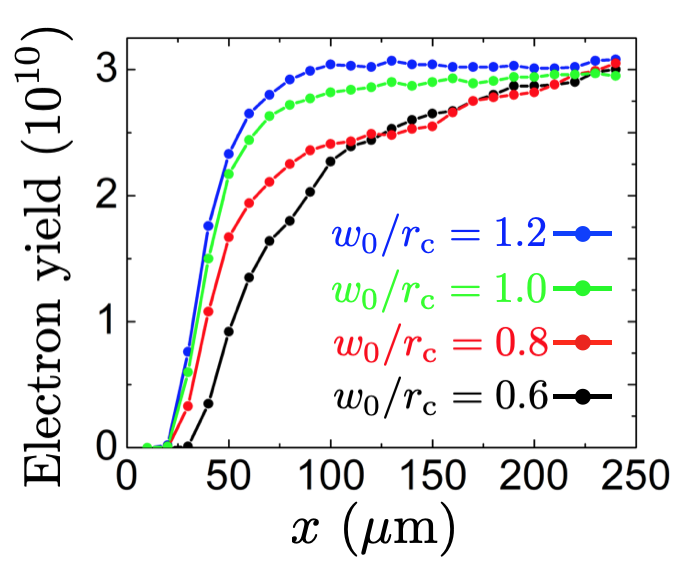}\caption{\label{f2} (Color online) The electron yield vs propagation distance, for different $w_0/r_{\rm{c}}$ ratios.}
\vspace{-1pt}
\end{figure}

Interestingly, despite different injection processes, 
all cases result in similar beam charges for sufficiently long MPW. This is because in order to be injected into the channel, the electrons percolating through the laser nodes must also overcome the electrostatic potential barrier near the MPW wall, which leads to saturation. The charge injected at the entrance suppresses the injection inside the channel.
In the end, the maximum charge separation on the wall will be just sufficient to prevent the most energetic counterstreaming electrons inside the MPW from escaping. The energy of these electrons is determined by the fundamental waveguide mode, which is the same for all cases.

To estimate the energy of the electrons we use momentum conservation.  Note that the ion response time is typically longer than the laser duration, therefore the photon momenta is first transferred to electrons. The number of plasma electrons  streaming towards the MPW inner surface (counterstream due to charge separation) per unit time is $N_{\rm{e}}  \approx n_{\rm{c}}\pi r_{\rm{c}}^2\beta^{'}ck_x/k_{\rm{T}}$, where $k_{\rm{T}}$ and $k_x$ are the transverse and longitudinal wavenumber in the MPW, $\beta^{'}$ is the radial velocity of counterstreaming electrons normalised by $c$. These electrons are reflected back on the plasma-vacuum interface due to the interaction with the photons ($N_{\gamma}$ per unit time). We assume the number of electrons percolating through the laser nodes as well as the number of the photons absorbed are negligibly small during this process. According to momentum conservation $2N_{\gamma}\hbar k_{\rm{T}} = N_{\rm{e}}(\gamma\beta_r+\gamma^\prime\beta^\prime)m_{\rm{e}}c$, where $\gamma^\prime = (1-\beta^{\prime 2})^{-1}$, $\gamma$ and $\beta_r$ are the relativistic gamma factor and the normalised radial velocity of the electrons that being pushed back (after the interaction). 

Here we are only interested in the maximum counterstreaming electron energy that can be achieved. Substituting $\beta_r \approx \beta^{'}$ due to quasi-neutrality, we find that when the longitudinal velocity of the surface electrons vanishes after interaction, $\gamma^\prime$ reaches its maximum,
\begin{equation}
\gamma_{\rm{m}}^\prime \approx \frac{\Gamma+\sqrt{\Gamma^2+4}}{2}
\label{eq1}
\end{equation}
{\noindent}where $\Gamma \equiv (x_1^2a_{\rm{m}}^2)/(k_0^2r_{\rm{c}}^2)$, $k_0 = \sqrt{k_x^2+k_{\rm{T}}^2}$, and $a_{\rm{m}}$ is the normalised intensity of the waveguide mode. For $w_0\geq r_{\rm{c}}$, $a_{\rm{m}} \approx a_0$, and for $w_0<r_{\rm{c}}$, $a_{\rm{m}} = a_0w_0/r_{\rm{c}}$. We have assumed the radius of MPW is sufficiently  large ($k_{\rm{T}}\ll k_0$), and only the fundamental mode exists inside the MPW, so that $k_{\rm{T}} = x_1/r_{\rm{c}}$ and $x_1 = 2.4$ is the first root of eigenvalue equation \cite{Shen1991}. 

The electrostatic field near the MPW wall can be estimated using Gauss's law, $E_{\rm{C}} = 2Q/r_{\rm{c}}c\tau_0$, where $Q$ is the charge that is lost from the wall (i.e.~injected into the channel). Further injection can only happen when the kinetic energy of conterstreaming electrons overcomes the electrostatic potential within the skin layer, i.e.~$(\gamma_{\rm{m}}^\prime-1)m_{\rm{e}}c^2\approx \sqrt{\gamma_{\rm{m}}^\prime}eE_{\rm{C}}c/\omega_0$, which yields the saturation charge,
\begin{equation} 
Q \approx \frac{(\gamma_{\rm{m}}^\prime-1)}{\sqrt{\gamma_{\rm{m}}^\prime}}\frac{k_0r_{\rm{c}}}{2}\frac{m_{\rm{e}}c^3}{e}\tau_0.
\label{eq2}
\end{equation}

\begin{figure}[!b]
\centering
\includegraphics[width=8.5cm]{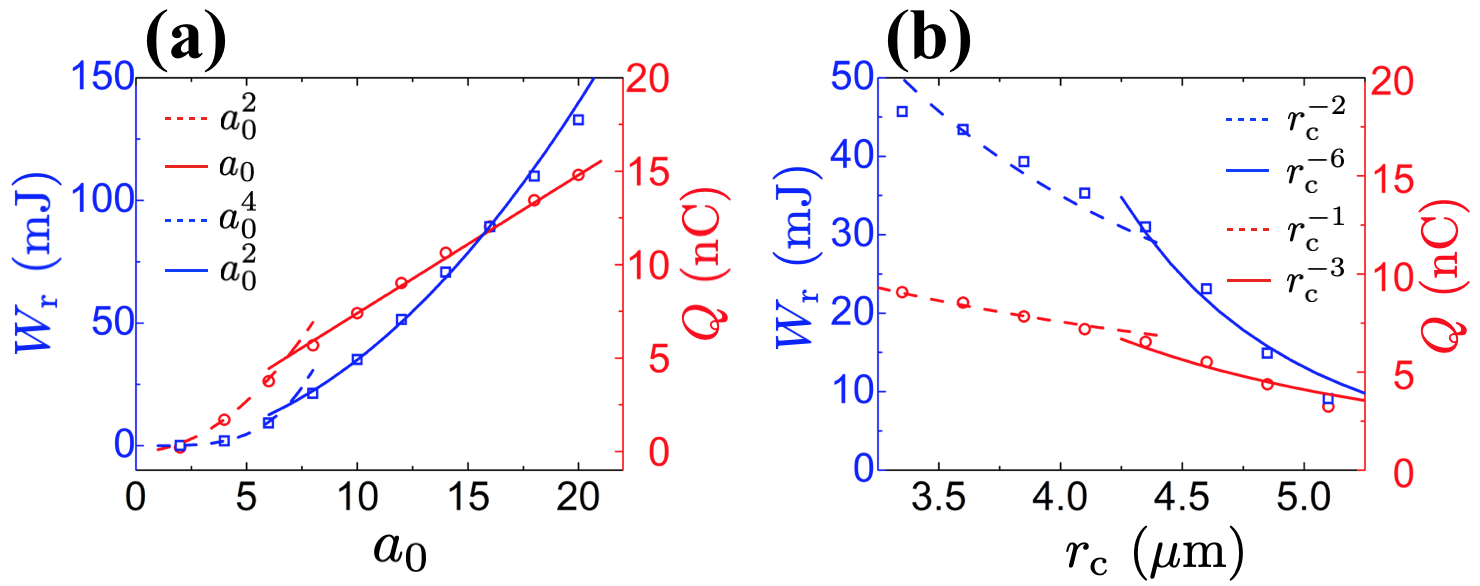}\caption{\label{f2} (Color online) The electron beam charge (red) and the THz radiation energy (blue) evolution with varying (a) laser intensity (with $w_0 = 4~\mu m$ and $r_{\rm{c}} = 3.35~\mu m$) and (b) effective MPW radii (with $a_0 = 10$ and $w_0 = 4~\mu m$). Open markers are 3D PIC simulation results (circles and squares represent the electron charge and THz energy respectively), and the solid and dashed curves are fittings suggested by Eq.~(2).} 
\vspace{-1pt}
\end{figure}

As an order-of-magnitude estimate, for a micro-sized channel, $k_0r_{\rm{c}}$ is typically around unity. This means that a $10^{20}$ W/cm$^{2}$, 50-fs laser system could produce 10~nC electron beams, which agrees with simulations. From Eq.~(2) the scaling of the charge with the normalised laser intensity can be estimated:  for weakly-relativistic cases ($\Gamma \ll 1$) $Q\propto a_{\rm{m}}^2/r_{\rm{c}}$, while for strongly-relativistic cases ($\Gamma \gg 1$) $Q\propto a_{\rm{m}}$. Note, that the energy of CDR scales as $W_{\rm{r}} \propto Q^2$, it is therefore important to confirm these scalings by 3D PIC simulations to guide future experiments.  In the analysis above we neglected the azimuthal dependence, which is strictly valid only for circular polarization. However, 3D PIC simulations presented in Fig.~4 show that  the obtained scalings are valid also for linear polarization. 

In Fig.~4(a), we plot the electron charge produced by the MPW and the total THz energy (below 60 THz) as functions of $a_0$, where $r_{\rm{c}} = 3.35~\mu$m and $w_0 = 4~\mu$m are fixed, and the MPW length is $L = 30~\mu$m. The parameters are the same as in Fig.~1 unless otherwise stated. It is shown that the charge increases quadratically with the normalised laser intensity ($Q\propto a_0^2$) when $a_0$ is small, and the scaling becomes linear ($Q\propto a_0$) for $a_0 > 8$, where $\Gamma$ exceeds unity. In addition, the simulation results indicate that the THz energy can be fitted by $W_{\rm{r}}\propto a_0^4$ in the weakly-relativistic regime, where the transition efficiency increases linearly with the intensity. In the strongly-relativistic regime, the THz energy can be fitted by $W_{\rm{r}}\propto a_0^2$, where the conversion efficiency saturates at $\sim 1\%$. Note, that this also demonstrates that the radiation is coherent ($W_{\rm{r}}\propto Q^2$). Our numerical results suggest that TW-class, 100-mJ-strong THz emission can be produced by a 10-J/250-TW laser system, which is within reach of the existing laser facilities.

In Fig.~4(b), we consider the effects of varying the MPW radius variation when the laser parameters are fixed ($a_0 = 10$ and $w_0 = 4~\mu$m), and the MPW length is extended to $L = 120~\mu$m to ensure sufficient distance for injection. In this case, Eq.~(2) leads to $Q\propto r_{\rm{c}}^{-1}$ in the strongly-relativistic regime, and $Q\propto r_{\rm{c}}^{-3}$ in the weakly-relativistic regime, which agrees with our simulations. Since the radiation is coherent, it results in a quenching effect: the radiation energy drops dramatically ($W_{\rm{r}}\propto r_{\rm{c}}^{-6}$) as the effective radius exceeds a threshold near $\Gamma\sim 1$. This is verified by a sharp decrease of the THz energy at the separatrix of the two regimes around $r_{\rm{c}}\approx 4.3~\mu$m, when the counterstreaming electrons become weakly-relativistic (i.e.~$\Gamma \approx 0.7$).

Finally, we note that Eq.~(2) does not consider the effects of high-order waveguide modes (which give a higher transverse light pressure) and strong diffraction at the entrance (so that the charge produced at the entrance may already exceed the saturation limit suggested by Eq.~(2), especially in large MPW). In fact, the value from Eq.~(2) should be treated as the minimum charge that can be produced by laser-MPW interaction, as the interaction between the lowest-order mode and MPW is the weakest. A detailed study of these effects is left for future work.


In conclusion, we proposed a scheme to generate relativistic isolated THz pulses based on the interaction of a laser pulse with a micro-plasma-waveguide. 3D PIC simulations show the energetic electron beam with a few tens of nC charge can be produced inside the channel. As the beam exits the waveguide, a substantial part of the electron energy is transferred to an intense THz emission through coherent diffraction radiation. We demonstrated with 3D PIC simulations that the overall efficiency reaches 1$\%$, the radiation power 1 TW and the energy 100 mJ. We obtain scaling laws for THz generation energy in different regimes that are characterised by the maximum gamma factor of the counterstreaming electrons induced by the fundamental mode. The proposed scheme can be easily extended to other frequency ranges by varying the driving laser duration, allowing the generation of radiation from infrared to sub-THz range with relativistic intensities. This opens a new avenue towards high-power light matter interaction beyond the state of the art.

\begin{acknowledgments}
The authors acknowledge fruitful discussions with I Thiele, S Newton, I Pusztai, E Siminos, and J Ferri. This work is supported by the Olle Engqvist Foundation, the Knut and Alice Wallenberg Foundation and the European Research Council (ERC-2014-CoG grant 647121). Simulations were performed on resources at Chalmers Centre for Computational Science and Engineering (C3SE) provided by the Swedish National Infrastructure for Computing (SNIC).
\end{acknowledgments}


\begin{thebibliography}{99}\suppressfloats
\bibitem{Tonouchi2007} {M. Tonouchi}, Nat. Photonics \textbf{1}, 97 (2007).
\bibitem{Siegel2004} {P. H. Siegel}, IEEE Trans. Microwave Theory Tech. \textbf{52}, 2438 (2004).
\bibitem{Cole2001} {B. E. Cole, J. B. Williams, B. T. King, M. S. Shervin, and C. R. Stanley}, Nature \textbf{410}, 60 (2001).
\bibitem{Hoffmann2011} {M. C. Hoffmann and J. A. F\"ul\"op}, J. Phys. D \textbf{44}, 083001 (2011).
\bibitem{Kampfrath2013} {T. Kampfrath, K. Tanaka, and K. Nelson}, Nat. Photonics \textbf{7}, 680 (2013).
\bibitem{Oh2014} {T. I. Oh, Y. J. Yoo, Y. S. You, and K. Y. Kim}, Appl. Rev. Lett. \textbf{112}, 213901 (2014).
\bibitem{Hirori2011} {H. Hirori, A. Doi, F. Blanchard, and K. Tanaka}, Appl. Phys. Lett. \textbf{98}, 091106 (2011).
\bibitem{Fulop2014} {J. A. F\"ul\"op, Z. Ollmann, Cs. Lombosi, C. Skrobol, S. Klingebiel, L. P\'alfalvi, F. Krausz, S. Karsch, and J. Hebling}, Opt. Express \textbf{22}, 20155 (2014).
\bibitem{Vicario2014} {C. Vicario, B. Monoszlai, and C. P. Hauri}, Phys. Rev. Lett. \textbf{112}, 213901 (2014).
\bibitem{Sheng2005} {Z. M. Sheng, K. Mima, J. Zhang, and H. Sanuki}, Phys. Rev. Lett. \textbf{94}, 095003 (2005).
\bibitem{Gopal2013} {A. Gopal, S. Herzer, A. Schmidt, P. Singh, A. Reinhard, W. Ziegler, D. Br\"ommel, A. Karmakar, P. Gibbon, U. Dillner, T. May, H. G. Meyer, and G. G. Paulus}, Phys. Rev. Lett. \textbf{111}, 074802 (2013).
\bibitem{Li2012} {Y. T. Li, C. Li, M. L. Zhou, W. M. Wang, F. Du, W. J. Ding, X. X. Lin, F. Liu, Z. M. Sheng, X. Y. Peng, L. M. Chen, J. L. Ma, X. Lu, Z. H. Wang, Z. Y. Wei, and J. Zhang}, Appl. Phys. Lett. \textbf{110}, 254101 (2012).
\bibitem{Liao2016} {G. Q. Liao, Y. T. Li, Y. H. Zhang, H. Liu, X. L. Ge, S. Yang, W. Q. Wei, X. H. Yuan, Y. Q. Deng, B. J. Zhu, Z. Zhang, W. M. Wang, Z. M. Sheng, L. M. Chen, X. Lin, J. L. Ma, X. Wang and J. Zhang}, Phys. Rev. Lett. \textbf{116}, 205003 (2016).
\bibitem{Tian2017} {Y. Tian, J. S. Liu, Y. F. Bai, S. Y. Zhou, H. Y. Sun, W. W. Liu, J. Y. Zhao,
R. X. Li, and Zhizhan Xu}, Nat. Photonics \textbf{11}, 242 (2017).
\bibitem{Chen2015} {Z. Y. Chen and A. Pukhov}, Phys. Plasmas \textbf{22}, 103105 (2015).
\bibitem{Herzer2018} {S. Herzer, A. Woldegeorgis, J. Polz, A. Reinhard, M. Almassarani, B. Beleites, F. Ronneberger, R. Grosse, G. G. Paulus, U. H\"ubner, T. May and A. Gopal}, New J. Phys. \textbf{20}, 063019 (2018).
\bibitem{Liao2019} {G. Q. Liao, Y. T. Li, H. Liu, G. G. Scott, D. Neely, Y. H. Zhang, B. J. Zhu, Z. Zhang, C. Armstrong, E. Zemaityte, P. Bradford, P. G. Huggard, D. R. Rusby, P. McKenna, C. M. Brenner, N. C. Woolsey, W. M. Wang, Z. M. Sheng and J. Zhang}, PNAS \textbf{116}, 3994 (2019).
\bibitem{Thiele2019}{I. Thiele, E. Simios, and T. F\"ul\"op},  Phys. Rev. Lett. \textbf{122}, 104803 (2019).
\bibitem{Carr2002} {G. L. Carr, M. C. Martin, W. R. McKinney, K. Jordan, G. R. Neil, and G. P. Williams}, Nature \textbf{420}, 153 (2002).
\bibitem{Ginzburg1982} {V. L. Ginzburg}, Phys. Scr. \textbf{T2/1}, 182 (1982).
\bibitem{Happek1991} {U. Happek, A. J. Sievers, and E. B. Blum}, Phys. Rev. Lett. \textbf{67}, 2962 (1991).
\bibitem{Dnestrovskii1959} {Yu. N. Dnestrovskii and D. P. Kostomarov}, Dokl. Akad. Nauk \textbf{124}, 792 (1959) [Sov. Phys. Dokl. \textbf{4}, 132 (1959)]; \textbf{124}, 1026 (1959) [\textbf{4}, 158 (1959)].
\bibitem{Shibata1995} {Y. Shibata, S. Hasebe, K. Ishi, T. Takahashi, T. Ohsaka, and M. Ikezawa}, Phys. Rev. E \textbf{52}, 6787 (1995).
\bibitem{Schroeder2004} {C. B. Schroeder, E. Esarey, J. Tilborg, and W. P. Leemans}, Phys. Rev. E \textbf{69}, 016501 (2004).
\bibitem{Wu2013} {Z. Wu, A. S. Fisher, J. Goodfellow, M. Fushs, D. Darancinang, M. Hogan, H. Loos, and A. Lindenberg}, Rev. Sci. Instrum. \textbf{84}, 022701 (2013).
\bibitem{Leemans2014} {W. P. Leemans, A. J. Gonsalves, H. S. Mao, K. Nakajima, C. Benedetti, C. B. Schroeder, Cs. T\'oth, J. Daniels, D. E. Mittelberger, S. S. Bulanov, J. L. Vay, C. G. R. Geddes, and E. Esarey}, Phys. Rev. Lett. \textbf{113}, 245002 (2014).
\bibitem{Leemans2002} {W. P. Leemans, P. Catravas, E. Esarey, C. G. R. Geddes, C. Toth, R. Trines, C. B. Schroeder, B. A. Shadwick, J. van Tilborg, and J. Faure}, Phys. Rev. Lett. \textbf{89}, 174802 (2002).
\bibitem{Gahn1999} {C. Gahn, G. D. Tsakiris, A. Pukhov, J. Meyer-ter-Vehn, G. Pretzler, P. Thirolf, D. Habs, and K. J. Witte}, Phys. Rev. Lett. \textbf{83}, 4772 (1999).
\bibitem{Bulanov1994} {S. V. Bulanov, F. F. Kamenets, F. Pegoraro, and A. M. Pukhov}, Phys. Rev. A \textbf{195}, 84 (1994).
\bibitem{Yi2016_1} {L. Q. Yi, A. Pukhov, P. Luu-Thanh, and B. F. Shen}, Phys. Rev. Lett. \textbf{116}, 115001 (2016).
\bibitem{Yi2016_2} {L. Q. Yi, A. Pukhov, and B. F. Shen}, Phys. Plasmas \textbf{23}, 073110 (2016).
\bibitem{Yi2016_3} {L. Q. Yi, A. Pukhov, and B. F. Shen}, Sci. Rep. \textbf{6}, 28147 (2016).
\bibitem{Gong2019} {Z. Gong, A. P. L. Robinson, X. Q. Yan, and A. V. Arefiev}, Plasma Phys. Control Fusion \textbf{61}, 035012 (2019).
\bibitem{Naumova2004} {N. Naumova, I. Sokolov, J. Nees, A. Maksimchuk, V. Yanovsky and G. Mourou}, Phys. Rev. Lett. \textbf{93}, 195003 (2004).
\bibitem{Arber2015} {T. D. Arber, K. Bennett, C. S. Brady, A. Lawrence-Douglas, M. G. Ramsay, N. J. Sircombe, P. Gillies, R. G. Evans, H. Schmitz, A. R. Bell, and C. P. Ridgers}, Plasma Phys. Control. Fusion \textbf{57}, 113001 (2015).
\bibitem{Cowan2013} {B. M. Cowan, D. L. Bruhwiler, J. R. Cary, and E. Cormier-Michel}, Phys. Rev. Accel. Beams \textbf{16}, 041303 (2013).
\bibitem{Carron2000} {N. J. Carron}, Prog. Electromagnetics Res. \textbf{28}, 147 (2000).
\bibitem{Shen1991} {H. M. Shen}, J. Appl. Phys. \textbf{69}, 6827 (1991).

\end{thebibliography}
\end{document}